# *Ab Initio* Evaluation of Plasmon Lifetimes in the Alkali Metals

*Wei Ku and Adolfo G. Eguiluz*

Department of Physics and Astronomy, The University of Tennessee, Knoxville, TN 37996-1200
and Solid State Division, Oak Ridge National Laboratory, Oak Ridge, TN 37831-6030

**Abstract**

The anomalous plasmon linewidth dispersion (PLD) measured in K by vom Felde, Sprösser-Prou, and Fink (Phys. Rev. B **40**, 10181 (1989)), has been attributed to strong dynamical electron-electron correlations. On the basis of *ab initio* response calculations, and detailed comparison with experiment, we show that the PLD of K is, in fact, dominated by decay into particle-hole excitations involving empty states of *d*-symmetry. For Li, we shed new light on the physics of the PLD. Our all-electron results illustrate the importance of *ab initio* methods for the study of electronic excitations.

The lifetime of the electronic excitations contains valuable signatures of the physics of correlated electrons in solids. However, the identification of the mechanisms responsible for the measured linewidths is often made on the basis of simplified models which may lack some of the key ingredients of the decay process as it is realized in nature. As a consequence, it becomes difficult to assess in a realistic way the importance of the effects of the one-electron band structure relative to those of the many-electron correlations.

A case in point is provided by landmark experiments on the electronic excitations in the alkali metals by vom Felde, Sprösser-Prou, and Fink [1]. Their high-resolution electron-energy-loss spectroscopy (EELS) data revealed remarkable anomalies in the plasmon spectra of these systems [1,2]. In particular, fundamental questions were raised regarding the mechanism behind the plasmon linewidth dispersion (PLD) of K, which was found to be *positive*, for all wave vectors [3]. This behavior is in conflict with the predictions of the best available theory of plasmon damping due to interband transitions [4,5], which yields a *negative* PLD.

The theoretical approach in question, due to Sturm [4] and Sturm and Oliveira [5], is based on the evaluation of the dielectric function to second order in the pseudopotential, which is assumed to be weak. A phenomenological inclusion of correlations involving the decay of the plasmon into two electron-hole pair states [6] yields a (small) positive slope for the PLD curve [5]; however, the quantitative discrepancy with experiment is still enormous [1,2].

Now K is generally considered to be an almost ideal realization of the jellium model of interacting electrons. Moreover, the EELS data for K show no fine structure due to single-particle transitions [2]. On this basis, and in view of the above conflict between theory and experiment, it was concluded [1] that the PLD of K must be controlled by strong correlation effects, involving higher order processes in the electron-electron interaction.

By contrast, in the case of Li the perturbation-theoretic approach [5] appears to be successful —to judge by its overall qualitative agreement with early EELS data [7,8] and purported support from *ab initio* calculations [9] (which, however, refer to rather large wave vectors). This is surprising, since the gap at the (110) zone boundary (*N*-point) is comparable with the occupied bandwidth. Thus the validity of a low-order perturbation treatment of the Li lattice potential is not obvious. Similarly, it is intriguing that the perturbation theory may fail for K, of all systems, since in this case the gap at the *N*-point *is* small.

In this Letter we report *ab initio* calculations which provide a consistent picture of the PLD in both systems. We show that the key role in understanding the vom Felde et al. [1,2] data for K is played by one-particle states of *d*-symmetry. This damping mechanism, which is entirely absent from previous theory [5], accounts for the observed PLD quite accurately. Dynamical correlations are thus ruled out as a significant damping process. For Li, our PLD curve differs appreciably from the perturbation-theoretic result; since empty states of *d*-character are not involved, this difference is a signature of the strong Li crystal potential.

Our *ab initio* method is based on the evaluation of the dynamical density-response function for imaginary frequencies, followed by analytic continuation to the real axis. This approach allows us to extract the natural plasmon linewidth accurately. Existing real-axis algorithms [9,10,11] do not lend themselves easily to linewidth determination.

In the framework of time-dependent density-functional theory (TD-DFT) [12], the retarded density-response function $\chi(\vec{x},\vec{x}';\omega)$ obeys the exact integral equation

$$\chi = \chi^{(0)} + \chi^{(0)}(v + f_{xc})\chi \ , \tag{1}$$

where $\chi^{(0)}(\vec{x},\vec{x}';\omega)$ is the response function for non-interacting Kohn-Sham electrons, $v(\vec{x}-\vec{x}')$ is the bare Coulomb interaction, and the vertex $f_{xc}$ accounts for exchange and correlation effects. For a periodic crystal the Fourier transform of the polarizability is given by

$$\chi^{(0)}_{\vec{G},\vec{G}'}(\vec{q};\omega) = \frac{1}{V}\sum_{\vec{k}}^{BZ}\sum_{n,n'}\frac{f_{\vec{k},n} - f_{\vec{k}+\vec{q},n'}}{E_{\vec{k},n} - E_{\vec{k}+\vec{q},n'} + \hbar(\omega + i\eta)}\left\langle \vec{k},n \middle| e^{-i(\vec{q}+\vec{G})\cdot\hat{\vec{x}}} \middle| \vec{k}+\vec{q},n' \right\rangle$$
$$\times \quad \left\langle \vec{k}+\vec{q},n' \middle| e^{i(\vec{q}+\vec{G}')\cdot\hat{\vec{x}}} \middle| \vec{k},n \right\rangle \ , \tag{2}$$

where $\vec{G}$ is a vector of the reciprocal lattice, $n$ is a band index, the wave vectors $\vec{k}$ and $\vec{q}$ are in the first Brillouin zone (BZ), and $V$ is the normalization volume. In this representation Eq. (1) is turned into a matrix equation which we solve numerically; the plasmon loss corresponds to a well-defined peak in $\mathrm{Im}\,\chi_{\vec{G}=0,\vec{G}'=0}(\vec{q};\omega)$ for a given $\vec{q}$. The off-diagonal matrix elements in Eq. (2) incorporate the "crystal-local fields" in the response function; these effects turn out to play a minor role in the present calculations —thus, they will not be mentioned any further.

We have developed a code to evaluate Eq. (2) in the local-density approximation (LDA) [13], starting from the knowledge of the band structure and wave functions [14] in the full-potential linearized augmented plane wave (LAPW) method [15]. The local-orbital extension of the LAPW method [15] permits an accurate treatment of the contribution to Eq. (2) from higher-lying core (or "semicore" ) states [16,17].

We aim at disentangling the effects of the band structure from those of dynamical many-body correlations. The TD-DFT framework [12] is well suited for this purpose, as it allows us to turn off the latter by setting $f_{xc} = 0$ —which we do. We compute $\chi^{(0)}$, and solve Eq. (1), for imaginary frequencies; the analytic continuation of $\chi$ to the real axis —a distance $\delta$ above it— is performed via Padé approximants [18] (note that the BZ is sampled without numerical broadening; $\eta \equiv 0$ in Eq. (2)). The power of our method is illustrated in Fig. (1), in which we compare the loss function obtained as just outlined [19] with its counterpart obtained via the more conventional evaluation of Eq. (2) for real $\omega$'s (a finite $\eta$ is now required) [9,10,11]. Figure 1 corresponds to a 16x16x16 Monkhorst-Pack mesh [20]; for the same mesh, as shown in the figure, the real-axis approach cannot resolve the linewidth. Note that our calculated loss peak is quite insensitive to a decrease of $\delta$ by an order of magnitude (in this regime, the numerical error is found to scale linearly with $\delta$). Since in the upper panel $2\delta$ is ~ 1/100 of the plasmon linewidth, an accurate value of the natural width is extracted [21].

In Fig. 2 we compare our calculated PLD for K with the EELS data of vom Felde et al. [1], together with the result of Ref. [5]. The left panel displays a well-converged PLD, obtained using a 20x20x20 $\vec{k}$-mesh and an energy cutoff of 20 eV in Eq. (2), corresponding to the inclusion of ~ 20 valence bands; the semicore excitations are also included [17]. The full-width at half-maximum of the plasmon peak, $\Delta E_{1/2}(\vec{q})$, is given relative to its extrapolated value for $\vec{q} = 0$. This convention [1] eliminates ambiguities in the absolute scale of $\Delta E_{1/2}(\vec{q})$, as the experimental linewidth contains contributions from, e.g., sample imperfections, which, however, do not yield a characteristic $\vec{q}$-dependence. We thus concentrate on the key issue — the linewidth *dispersion*. Clearly, our PLD curve is in excellent agreement with the EELS data [1]. Since our result does not include dynamical correlations ($f_{xc} = 0$), we conclude that the physics of the dispersion is not controlled by a many-body mechanism [1,22].

Insight into the physics of the PLD of K is gained on the right panel of Fig. 2, which shows results obtained upon restricting the number of bands kept in the evaluation of $\chi^{(0)}$. The bands relevant to the present argument [23] are shown in the inset, in which the shaded strip is the $\omega$-interval containing all the one-particle states that can couple directly to the plasmon [24]. Now *three* bands (thin solid lines) are needed in order to obtain a good dispersion curve for the plasmon *energy* [25]; keeping just these bands, our PLD curve (triangles) agrees well with the result of [5]. This is understandable, as the states kept are, for the most part, similar to the nearly-free-electron states entering the scheme of Ref. [5].

Next, we have the result (squares) which incorporates the contribution from three additional bands (thick solid lines in the inset). Note that it is just the lower edge of these bands that overlaps the shaded strip; this edge yields the upper half of the peak centered at ~ 3 eV in the density of states (DOS) of Fig. 3. The inclusion of these bands brings about a *qualitative* change in the PLD curve, which is now quite close to the EELS data [23]. Clearly, *these three bands provide the key decay channels for the plasmon of K.* It is significant that the bands in question are overwhelmingly of *d*-character —cf. DOS. The weak-pseudopotential scheme [5] replaces them with parabolic bands, which is a poor approximation. Additional insight into the PLD is gained by noting that [4]

$$\Delta E_{1/2}(\vec{q}) \approx \operatorname{Im}\varepsilon(\vec{q};\omega)\left(\partial \operatorname{Re}\varepsilon(\vec{q};\omega)/\partial\omega\right)^{-1}\Big|_{\omega=\omega_p(\vec{q})} , \qquad (3)$$

where $\varepsilon(\vec{q};\omega)=1-v(q)\chi^{(0)}(\vec{q};\omega)$ is the dielectric function (its $\vec{G}=0,\vec{G}'=0$ element). The first factor in Eq. (3) represents the availability of channels for decay into particle-hole pairs with energy $\hbar\omega_p(\vec{q})$ (given by the position of the peak in $\mathrm{Im}\,\chi(\vec{q};\omega)$); the second represents a "polarization effect" from interband transitions taking place at other $\omega$'s (e.g., semicore excitations, upper part of the *d*-bands discussed above, etc.). In Fig. 3 we show $\mathrm{Re}\,\varepsilon$, $\mathrm{Im}\,\varepsilon$, and $\mathrm{Im}\,\chi$, for the 3-band (dashes) and 6-band (thin solid lines) "simulations" of Fig. 2, for $\vec{q}=(3,3,0)(2\pi/(20a_0))$. Note that the inclusion of the second set of bands yields a large increase in $\mathrm{Im}\,\varepsilon$ —i.e., increased damping. Most importantly, decay into particle-hole pairs involving these *d*-bands yields a *positive* PLD; indeed, for $\vec{q}=(4,4,0)(2\pi/(20a_0))$ there is a marked enhancement of $\mathrm{Im}\,\varepsilon$ (thick solid line), with little change in the polarization factor. This can be understood as follows: Since $\omega_p(\vec{q})$ disperses upwards [1], a larger portion of the flat bands in the shaded strip of Fig. 2 becomes available with increasing $|\vec{q}|$ —and the damping process we have identified becomes more effective, in agreement with experiment [1].

The PLD of Li differs qualitatively from that of K, as illustrated by our results of Fig. 4 (solid circles). There is an initial positive dispersion for small *q*'s, which we have traced to matrix elements effects in $\chi^{(0)}$ [25], and a pronounced negative dispersion for larger *q*'s. This negative dispersion recognizes two origins, corresponding to the two factors in Eq. (3): First, unlike the case of K, there are fewer available decay channels ($\mathrm{Im}\,\varepsilon$ decreases) as $\omega$ increases in the vicinity of $\omega_p(\vec{q})$ (these channels are of *sp*-character). Next, the slope of $\mathrm{Re}\,\varepsilon$ increases with $|\vec{q}|$ for $\omega\cong\omega_p(\vec{q})$; this is a polarization effect of *sp*-bands lying above $\omega_p(\vec{q})$. Detailed analysis [25] shows —and this conclusion is new— that *the latter effect dominates the physics of the PLD in Li* (except at small *q*'s).

The empty triangles in Fig. 4 are the *ab initio* results of Karlsson and Aryasetiawan (who also solved Eq. (1) with $f_{xc} = 0$) [9]. Their data consist of just four points for relatively large $q$'s —the last two lie in the Landau-damping regime (which sets in for $|\vec{q}| \cong 0.6 k_F$, see Fig. 4). Given the non-trivial technical differences between both methods, it is pleasing that the results of [9] fit in well with our PLD curve. Our result, which treats the Li crystal potential exactly [26], differs appreciably from the perturbation-theoretic result of Sturm's [4]. Now for small $q$'s the plasmon loss is quite asymmetric (see inset), contrary to the assumption made in the scheme of Ref. [4]; this asymmetry is a direct consequence of the mechanism we just identified —the polarization effect symbolized by the second factor in Eq. (3). The empty circles in the left panel of Fig. 4 are the PLD obtained upon "symmetrizing" the line shape (dashes in the inset). With this result at hand we conclude that about half the error built into the perturbation treatment [4] is accounted for by the actual lineshape of the plasmon loss.

A comparison with available EELS data is presented on the right panel of Fig. 4. Our PLD curve is rather close to the data of Gibbons et al. [7] (squares). However their $q$-resolution (basically, the $\Delta q$ between consecutive squares) is not high enough for their data to have a bearing on details such as are present in our PLD curve. The data of Kloos (diamonds) [8] do contain a (small) positive slope for small $q$'s; however, the $\Delta q$ between their first two data points is quite large. It is puzzling that our linewidth is larger than the experimental one for small *q's*. It would be very useful to have data with high $\vec{q}$- and $\omega$- resolution available in this rather critical $\vec{q}$-domain [27]. We emphasize that, with the theoretical advance illustrated in Fig. 1, we are now in a position to compare in detail not just peak positions, but also actual lineshapes. Such comparison would provide a stringent test of our theory.

In summary, we have presented *ab initio* calculations of the PLD in K and Li. For K we have elucidated a fundamental issue raised by experiment [1]. We have shown that the EELS data of Ref. [1] can be understood naturally, and explained quantitatively, in terms of plasmon decay into single particle-hole pairs involving empty *d*-states. Consequently, dynamical correlations —the mechanism originally conjectured to be responsible for the anomalous PLD— can only have a small impact on the linewidth *dispersion* [28]. For Li we have obtained novel features of, and insight into, the PLD. Our findings showcase the importance of *ab initio* investigations of electron dynamics in condensed matter physics.

We thank Christian Halloy and the Joint Institute for Computational Science for training and technical support with the use of the University of Tennessee IBM SP2 computer. Thanks are also due to Wolf-Dieter Schöne for help with the Padé approximant technique, and to Ben Larson for critical comments on the manuscript. This work was supported by NSF Grant No. DMR-9634502 and the National Energy Research Supercomputer Center. ORNL is managed by Lockheed Martin Energy Research Corp. for the Division of Materials Sciences, U.S. DOE under contract DE-AC05-96OR2464.

28. It is straightforward to incorporate the effects of $f_{xc}$ in Eq. (1) if we invoke the LDA; this insertion leads to small changes in the results of Figs. 2 and 4 [25]. Multipair excitations would enter the theory via an $\omega$-dependent $f_{xc}$.

## FIGURE CAPTIONS

Fig. 1. Loss function for K for $\vec{q} = (1,1,0)\left(2\pi/\left(16a_0\right)\right)$, where $a_0 = 5.23\,\text{Å}$ is the lattice constant. Thick solid line: $\text{Im}\,\chi$ obtained from Eqs. (1) and (2) for imaginary $\omega$'s, followed by analytic continuation to a distance $\delta$ above the real-$\omega$ axis [19]. Thin solid line: $\text{Im}\,\chi$ obtained directly for real $\omega$'s, using a broadening parameter $\eta$ in Eq. (2). Top (bottom) panel corresponds to $\delta = \eta = 1\,\text{meV}$ ($10\,\text{meV}$).

Fig. 2. PLD for K. Comparison of our theoretical results with the EELS data of Ref. [1] (diamonds), and the theoretical results of Ref. [5] (solid line). Theory is for (1,1,0) propagation; the EELS data are for polycrystalline K. Left panel: Fully-converged calculation (solid circles); see text. Right panel: PLD's obtained upon keeping only 3 (triangles) and 6 (squares) valence bands in $\chi^{(0)}$ [23]. Inset: LDA band structure of K; the arrow indicates the value of $\omega_p(0)$ (see text and [24]).

Fig. 3. (a) Calculated DOS for K; total DOS and contributions to it from states of *s*,*p*, and *d*- symmetry [14]. (b) Real and imaginary parts of the dielectric function, and loss function $\text{Im}\,\chi$, for $\omega \approx \omega_p(\vec{q})$, for $\vec{q} = (3,3,0)\left(2\pi/\left(20a_0\right)\right)$. Results shown are for tests in which 3 (dashes) and 6 (thin solid lines) bands are kept in Eq. (2) (see Fig. 2). Thick solid line: $\text{Im}\,\varepsilon$ for the 6-band calculation for $\vec{q} = (4,4,0)\left(2\pi/\left(20a_0\right)\right)$.

Fig. 4. PLD for Li. Left panel: Comparison of our results (solid circles) with the theoretical PLD of [4] (solid line) and the ab initio results of [9] (empty triangles). The empty circles are our PLD for a symmetrized line shape, corresponding the dashes in the loss function shown in the inset for $\vec{q} = (2,2,0)\left(2\pi/\left(16a_0\right)\right)$; $|\vec{q}| \cong 0.28k_F$. Plasmon propagation is along the (1,1,0) direction. (Parameters used in our PLD are $\delta = 1\,\text{meV}$, 16x16x16 k-mesh, and 40 eV cutoff.) Right panel: Comparison with the EELS data of Ref. [7] and Ref. [8}. (Note: In the inset, the peak at ~ 3eV is a zone-boundary collective state [25].)

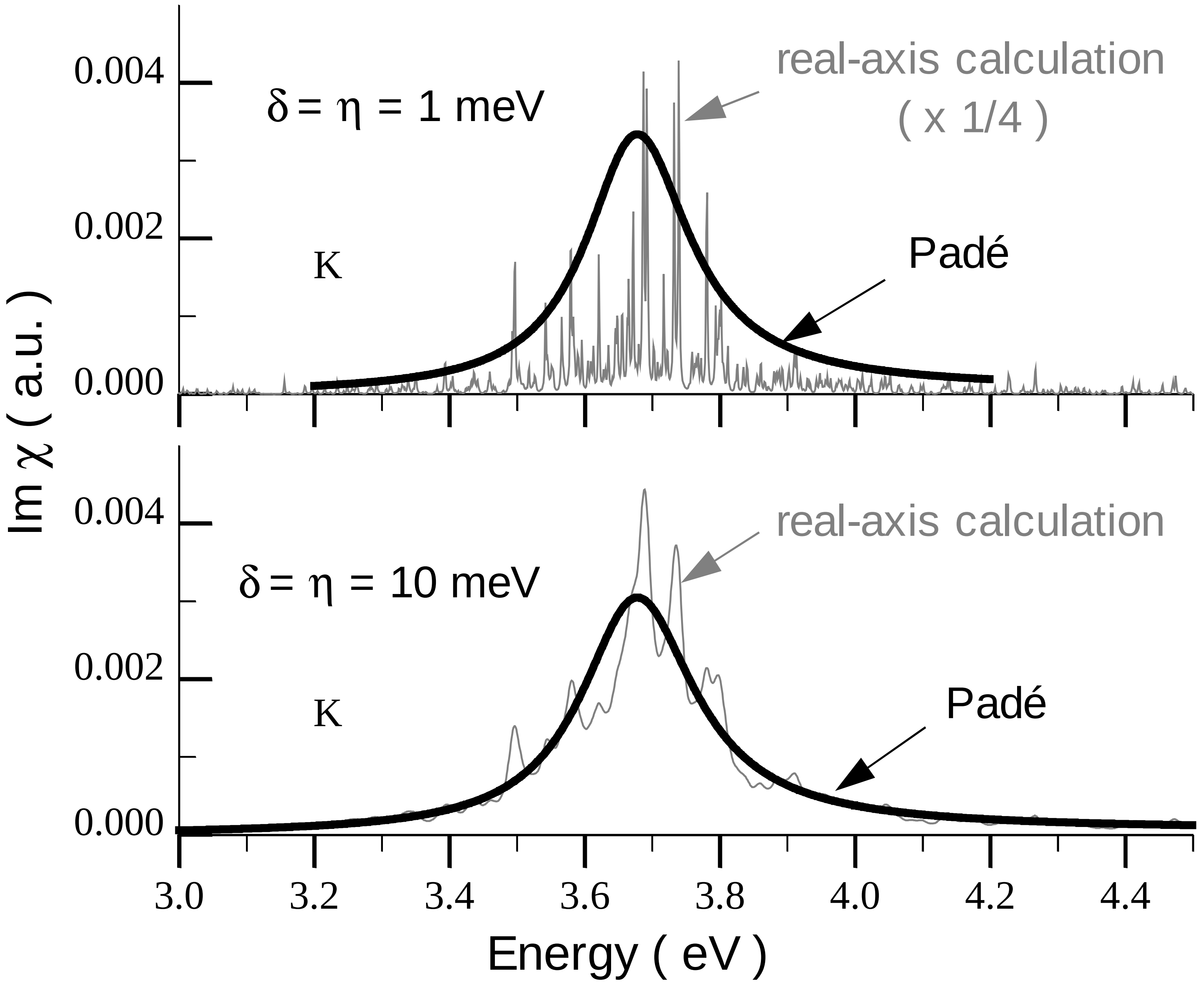

Fig. 1. Ku & Eguiluz

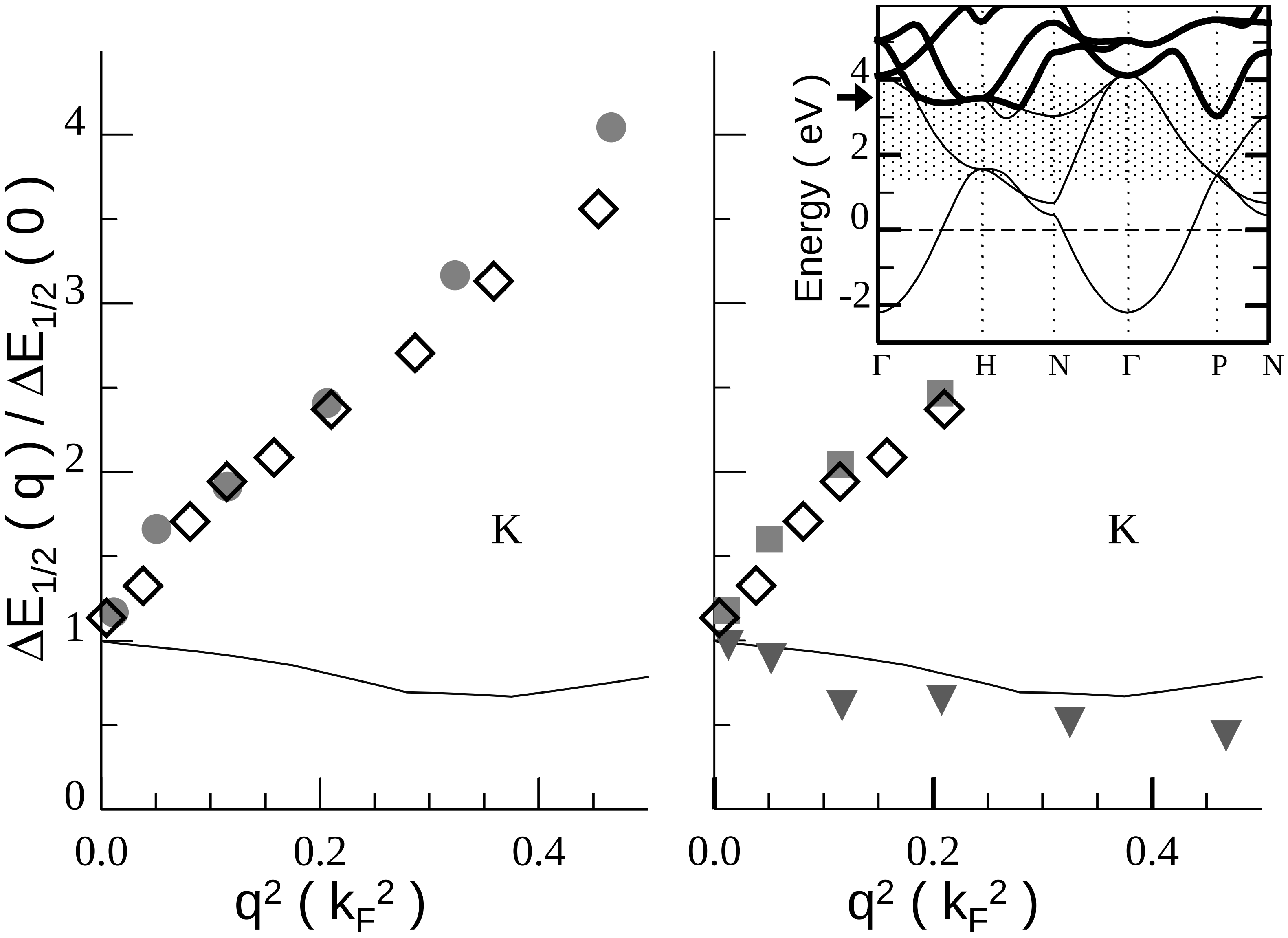

Fig. 2. Ku & Eguiluz

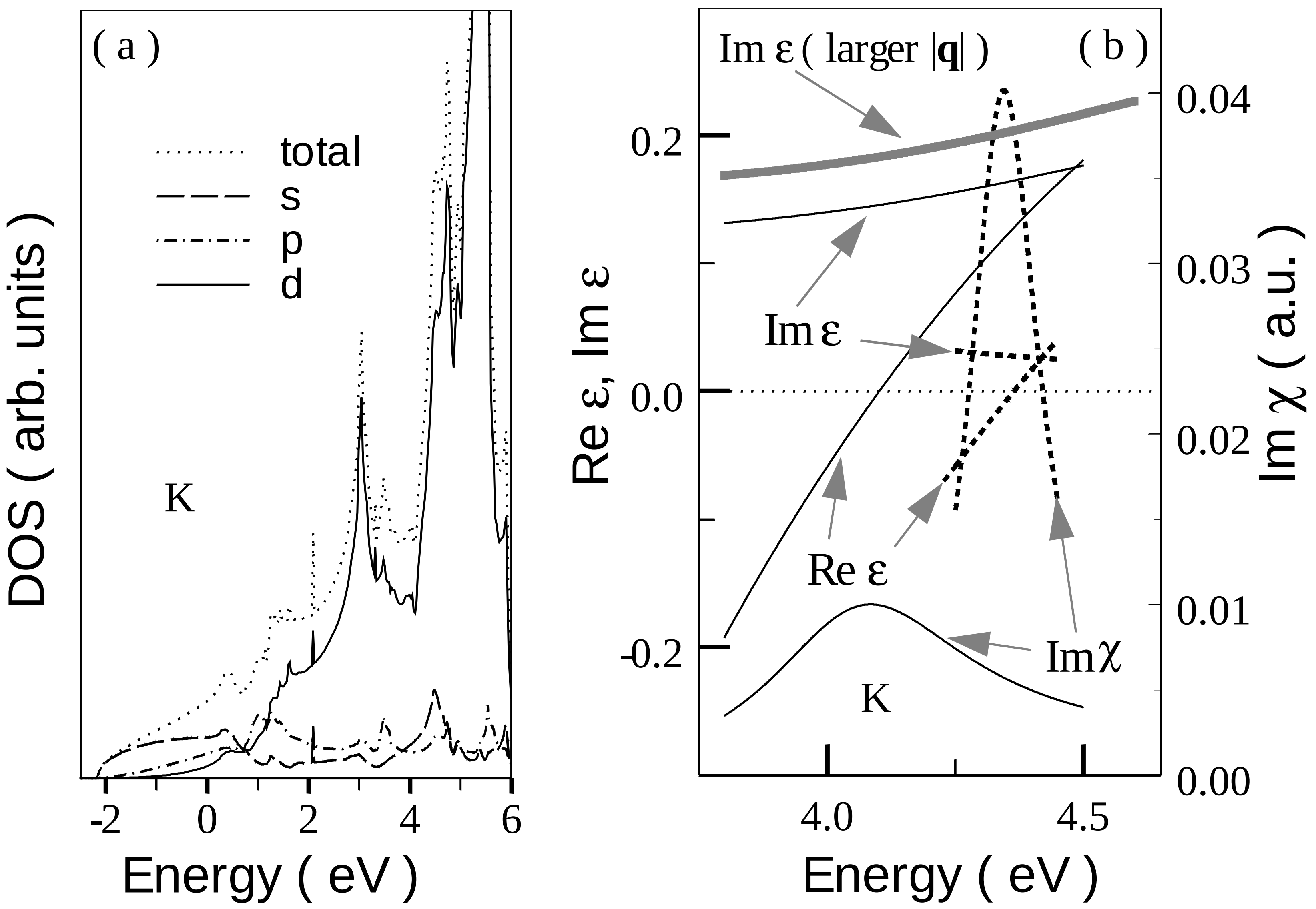

Fig. 3. Ku & Eguiluz

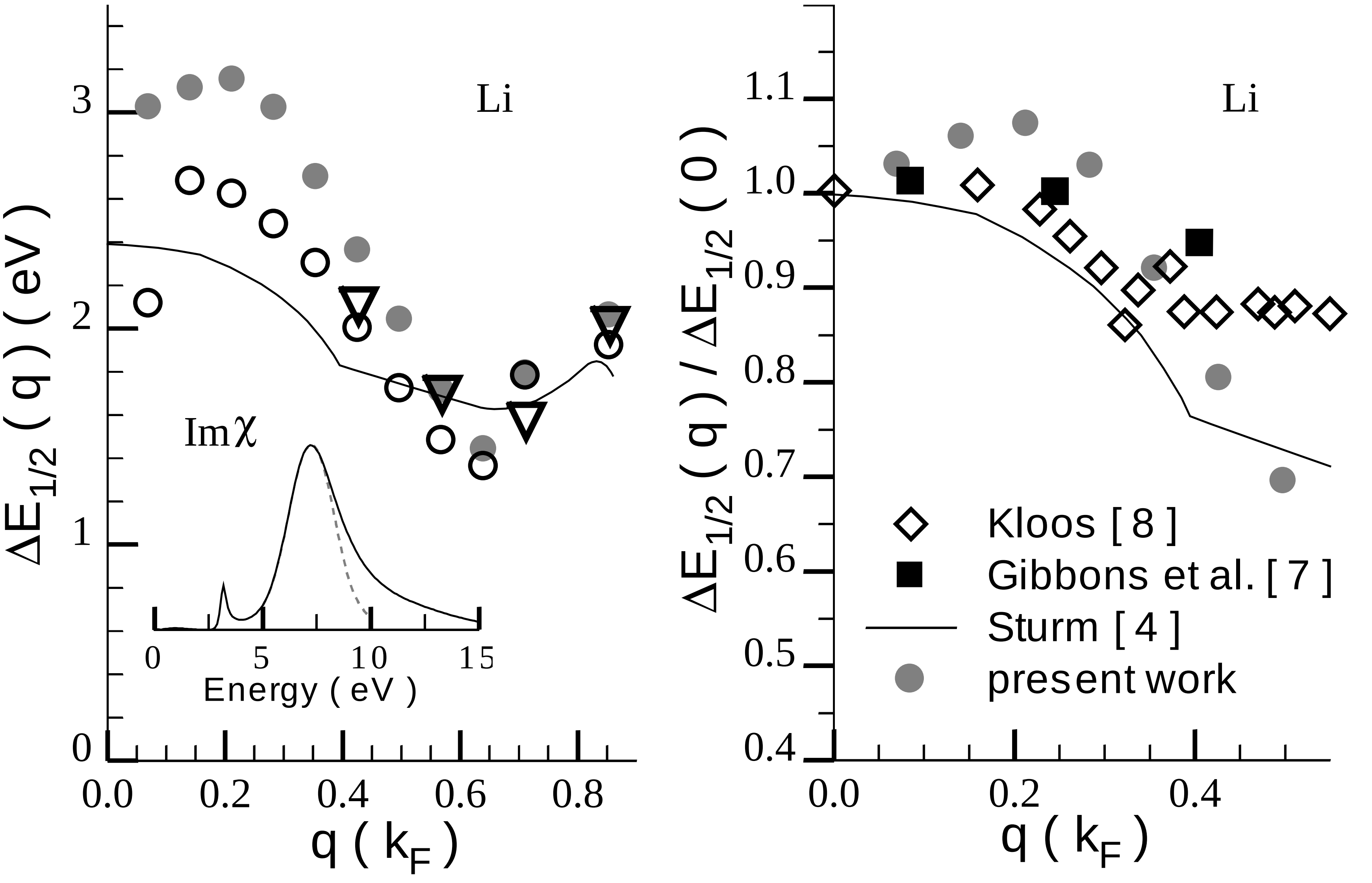

Fig. 4. Ku & Eguiluz